\documentclass[aip,amsmath,natbib,cite,mcite,reprint]{revtex4-1}
\usepackage[pdftex]{graphicx}
\begin{document}

\title{On the incorporation of cubic and hexagonal interfacial energy anisotropy 
in phase field models using higher order tensor terms} %Title of paper

\author{E.~S.~Nani}
\email{naniiitb777@gmail.com}
\author{M.~P.~Gururajan}
\email{gururajan.mp@gmail.com}
\thanks{Corresponding author}

\affiliation{Department of Metallurgical Engineering and Materials Science, Indian Institute of Technology Bombay, Powai, Mumbai 400076 INDIA} 

\date{\today}

\begin{abstract}

In this paper, we show how to incorporate cubic and hexagonal anisotropies 
in interfacial energies in phase field models; this incorporation is achieved by including 
upto sixth rank tensor terms in the free energy expansion, assuming that the free energy is 
only a function of coarse grained composition, its gradient, curvature and aberration. We 
derive the number of non-zero and independent components of these tensors. Further, by 
demanding that the resultant interfacial energy is positive definite for inclusion of each 
of the tensor terms individually, we identify the constraints 
imposed on the independent components of these tensors. The existing results in the invariant 
group theory literature can be used to simplify the process of construction of some 
(but not all) of the higher order tensors. Finally, we derive the relevant phase field 
evolution equations. 

\end{abstract}

\pacs{68.35.-p,68.35.Fx,61.50.Ah,02.70.-c,81.30.Hd,81.30.Mh} % insert suggested PACS numbers in braces on next line

\maketitle

\section{Introduction}

Phase field models are ideally suited, and hence are extensively used, to study microstructural 
evolution~\cite{ChenAnnRevMatSci,BoettingerEtAlAnnRevMatSci,Steinbach,Voorhees,LeuvenGang}.
Anisotropies play a crucial role in the formation and evolution of microstructures. 
The origins of anisotropy could be energetic (such as anisotropies in interfacial, elastic 
or magnetic energies) and/or kinetic (such as anisotropies in the attachment kinetics). 
Hence, a large number of phase field models have been developed to account for these
anisotropies: even though it is not possible to list all the phase field studies that deal with
anisotropies within the purview of this article, the following listing is fairly representative:
see, for interfacial anisotropy~\cite{AbiHaider,Ian,WangEtAl,MoelansEtAl,Lowengrub,Wheeler,McFaddenEtAl,Langer,BraunEtAl,
BraunEtAl2,CahnEtAl,HaxhimaliEtAl,QinBhadeshia1,QinBhadeshia2}; for elastic anisotropy~\cite{HeoSaswataChen,JinWangKhachaturyan,NiHeSoh,WangBanerjeeSuKhachaturyan,WangJinCuitinoKhachaturyan,
WangJinKhachaturyan}, for magnetocrystalline anisotropy~\cite{ZhangChen}, and, for anisotropy in attachment kinetics~\cite{Sekerka}.  

In a typical phase field model, the microstructure is described by order parameters and the
thermodynamic quantities (free energy or entropy) are represented as functionals in these order
parameters. The change in order parameters with time (and hence the microstructural evolution) 
is described in terms of the variational derivatives of the free energy with respect to these 
order parameters. Hence, it is natural that the energetic anisotropies are accounted in the 
phase field models through the free energies, while the kinetic anisotropies are accounted for 
through the relaxation parameters. 

The studies on the incorporation of kinetic anisotropies are relatively few while 
studies which incorporate the energetic anisotropies (be it interfacial, elastic or magnetic) 
are many. Almost all the phase field models that incorporate the anisotropies in the elastic 
(magnetic) energies do it the same way, namely, by including the anisotropy in the elastic  
(magnetic) energy term through the elastic moduli tensor (magnetic property tensor); however, 
sometimes, the interfacial energy anisotropy is incorporated without taking recourse to tensor 
terms explicitly: see, for example, Haxhimali et al~\cite{HaxhimaliEtAl}, and 
Qin and Bhadeshia~\cite{QinBhadeshia1,QinBhadeshia2}.

In this paper, we concentrate on incorporating the interfacial energy anisotropy by including
higher order (tensor) terms in the Taylor series expansion of the free energy. This is a well known 
method. In their classic paper, Cahn and Hilliard~\cite{Cahn-Hilliard} expanded the free energy
upto second rank terms (by including the gradients and curvatures of the local composition profile).
Such a second rank term cannot be used to capture cubic anisotropies in interfacial energy. Hence,
Abinandanan and Haider~\cite{AbiHaider} expanded the free energy upto fourth rank tensors 
(by including gradients, curvatures, aberrations and fourth derivatives of compositions). While cubic 
anisotropy is captured by these fourth rank tensors, for hexagonal systems, inclusion upto sixth rank 
tensor terms are essential. Thus, by expanding upto sixth rank tensors we can deal with both cubic 
and hexagonal crystal systems which are probably the most important ones for metals and alloys. 
Further, for isotropic systems as well as symmetries such as tetragonal, second rank tensor itself 
is sufficient. Hence, our aim in this paper is to extend the formulation of 
Abinandanan and Haider~\cite{AbiHaider} to include upto sixth rank tensors; this extension allows 
us to (a) include six fold anisotropy; and (b) make a cubic anisotropic system prefer $<110>$ 
directions over both $<100>$ and $<111>$ easily (which is harder to achieve by truncating the 
free energy expansion only upto the fourth rank tensor terms). 

We also derive the number of independent and non-zero components for each of the tensors as well as 
the constraints imposed on them. The derivation of the constraints (along with the number 
of independent components) is a key result. We believe that the identification of such constraints
play a crucial role in obtaining the anisotropy parameters either from experiments or 
by using other computational and/or simulation methods (such as molecular dynamics, 
Monte Carlo and/or first principles). The constraints 
derivation (under the given assumptions under which we derive them) also indicate that generic cross 
terms (for example, terms of the type that depend  both on gradient and curvature) are identically zero.

Finally, we show that the existing (and fairly well known) results in the group theory and invariant 
theory literature can help make the process of writing the free energy expansions rather straightforward; 
specifically, using the sixth rank tensors needed to incorporate hexagonal anisotropies as an example problem, 
we show how the existing group theoretical and invariant theory literature can be used as a recipe to write 
the required free energy expansions under certain restrictions, namely, that the free energy depends only 
on the local values of the order parameters, their gradients and their curvatures. However, it is possible 
that the free energy depends on higher derivatives; our formulation does include one such higher derivative, 
namely, the aberrancy. In such a case, we show how to derive the equivalent expressions through (laborius
but fairly straight-forward) calculations. 

The rest of this paper is organised as follows: in Section~\ref{Section2} of this paper, we describe our extended 
formulation to include upto sixth rank tensors; in Section~\ref{Section3}, using symmetry (intrinsic as well as 
crystalline) arguments we deduce the total number of independent and non-zero components for isotropic, cubic and 
hexagonal systems; in Section~\ref{Section4}, by demanding that the interfacial energy is always positive
we deduce the restrictions on the independent components for the three cases, namely,
isotropic, cubic  and hexagonal. In Section~\ref{Section5} of this
paper, we set down our recipe as to how, reading of terms from certain
tables in the group theoretical or invariant theory literature, one can write down the
free energy expansions in polynomial form. Finally, in Section~\ref{Section6}, we show the phase field 
evolution equations obtained from the given free energy. We conclude the paper with a brief summary of 
important results.

\section{Free energy including sixth rank tensor terms}\label{Section2}

In this paper, we consider systems whose microstructure can be described by a single, 
conserved order parameter, namely, (coarse-grained) composition. In the solidification 
literature, the thermodynamically consistent formalisms are based on the entropy functional. 
However, as shown by Plapp~\cite{Plapp}, there are distinct advantages to using free energy 
functionals even in the case of solidification.

Assuming that the free energy of the system depends {\bf \em only}
on the local values of the coarse-grained composition ($c$), gradients in composition ($c_i$, a vector),
curvature of the composition profile ($c_{ij}$, a second rank tensor), and aberrancy of the 
composition profile ($c_{ijk}$, a tensor of rank 3),
the total free energy $F$ of the system is written as
\begin{equation} \label{FreeEnergy}
F = \int_{V} f(c,c_i,c_{ij},c_{ijk}) dV
\end{equation}
where $f$ is the free energy density and $V$ is the volume of the system.

We assume that $f$ is a Taylor series
expansion, and restrict our expansion upto sixth rank tensor terms; further, 
without loss of generality, for the rest of this paper, 
we also assume that the systems
we are considering are all centro-symmetric. In centro-symmetric systems, 
as is well known, all the odd-ranked  tensors are identically zero (see Nye~\cite{Nye} for example). 
Hence, assuming Einstein summation
convention (of summation over repeated indices), the free energy can be expanded as follows:
\begin{equation}\label{FreeEnergyDensity}
f\left(c, c_i, c_{ij}, c_{ijk}\right) = f|_{0} + {\mathbf P}  + {\mathbf Q} + {\mathbf R} 
\end{equation}
where, the symbol $|_0$ represents (here and in the following equations) the value of the given quantity 
($f$, in this case) evaluated at $(c,0,0,0)$; that is, at a composition value of $c$ with the gradient, curvature and aberrancy being zero; $\mathbf P$ are terms involving second rank tensors, $\mathbf Q$ are terms involving fourth
rank tensors and $\mathbf R$ are terms involving sixth rank tensors. 

Specifically, there are two terms involving second rank tensors, namely,
\begin{equation}
{\mathbf P} = \kappa^I_{ij} c_i c_j+ \kappa^{II}_{ij} c_{ij}
\end{equation}
with
\begin{equation*}
\kappa^{I}_{ij}=\left.\frac{1}{2!}\,\frac {\partial^2 f}{\partial c_i \partial c_j }\right|_0, \hspace{20pt} \kappa^{II}_{ij}=\left.\frac{\partial f}{\partial c_{ij}}\right|_{0}.
\end{equation*}
As shown by Cahn and Hilliard~\cite{Cahn-Hilliard}, it is possible to use Gauss theorem and reduce the terms 
involving second rank tensors to one.

The number of terms involving fourth rank tensors are four, namely,
\begin{eqnarray}
{\mathbf Q} & = &  \beta^{I}_{ijkl} c_i c_j c_k c_l  +\beta^{II}_{ijkl} c_{ij} c_k c_l \nonumber \\
 & & + \beta^{III}_{ijkl} c_{ij} c_{kl} +\beta^{IV}_{ijkl} c_{ijk} c_l\nonumber   \\
\end{eqnarray}
with
\begin{equation*}
\beta^{I}_{ijkl}=\left.\frac{1}{4!}\,\frac{\partial^4 f}{\partial c_i \partial c_j \partial c_k \partial c_l}\right|_{0}, \hspace{5 pt}
\beta^{II}_{ijkl}=\left .\frac{1}{3!} \, \frac {\partial^3 f}{\partial c_{ij} \partial c_k \partial c_l}\right|_0, 
\end{equation*}
\begin{equation*}
\beta^{III}_{ijkl}=\left. \frac{1}{2!} \, \frac {\partial^2 f}{\partial c_{ij} \partial c_{kl} }\right|_0, \hspace{5 pt}
\beta^{IV}_{ijkl}=\left. \frac{1}{2!} \, \frac {\partial^2 f}{\partial c_{ijk} \partial c_{l} }\right|_0 \\
\end{equation*} 
As shown by Abinandanan and Haider~\cite{AbiHaider}, using Gauss theorem, the total number of fourth rank tensors can be reduced from four to three.

The number of terms involving sixth rank tensors are seven, namely,
\begin{eqnarray}
{\mathbf R} & = & \alpha^{I}_{ijklmn} c_i c_j c_k c_l c_m c_n +\alpha^{II}_{ijklmn} c_{ij} c_k c_l c_m c_n +\nonumber\\ 
&& \hspace{15pt}\alpha^{III}_{ijklmn} c_{ij} c_{kl} c_m c_n + \alpha^{IV}_{ijklmn} c_{ij} c_{kl} c_{mn}+\nonumber\\  
&&\hspace{15pt}\alpha^{V}_{ijklmn} c_{ijk} c_l c_m c_n+ \alpha^{VI}_{ijklmn} c_{ijk} c_{lm} c_n+ \nonumber\\ 
&&\hspace{15 pt}\alpha^{VII}_{ijklmn} c_{ijk} c_{lmn} \\ \nonumber   
\end{eqnarray}
with,
\begin{equation*}
\alpha^{I}_{ijklmn}=\left. \frac{1}{6!} \, \frac {\partial^6 f}{\partial c_{i} \partial c_{j} \partial c_{k} \partial c_{l} \partial c_{m} \partial c_{n} }\right|_0, 
\end{equation*}
\begin{equation*}
\alpha^{II}_{ijklmn}=\left. \frac{1}{5!} \, \frac {\partial^5 f}{\partial c_{ij} \partial c_{k}
\partial c_{l} \partial c_{m} \partial c_{n} }\right|_0,
\end{equation*}
\begin{equation*}
\alpha^{III}_{ijklmn}=\left. \frac{1}{4!} \, \frac {\partial^4 f}{\partial c_{ij} \partial c_{kl} \partial c_{m} \partial c_{n} }\right|_0,
\end{equation*}
\begin{equation*}
\alpha^{IV}_{ijklmn}=\left. \frac{1}{3!} \, \frac {\partial^3 f}{\partial c_{ij} \partial c_{kl}
\partial c_{mn} }\right|_0,
\end{equation*}
\begin{equation*}
\alpha^{V}_{ijklmn}=\left. \frac{1}{4!} \, \frac {\partial^4 f}{\partial c_{ijk} \partial c_{l} \partial c_{m} \partial c_{n} }\right|_0,
\end{equation*}
\begin{equation*}
\alpha^{VI}_{ijklmn}=\left. \frac{1}{3!} \, \frac {\partial^3 f}{\partial c_{ijl} \partial c_{lm}
\partial c_{n} }\right|_0,
\end{equation*}
\begin{equation*}
\alpha^{VII}_{ijklmn}=\left. \frac{1}{2!} \, \frac {\partial^2 f}{\partial c_{ijk} \partial c_{lmn} }\right|_0.
\end{equation*}

Of these seven tensors, the tensor term involving $\alpha^V_{ijklmn}$ can be reduced using Gauss theorem; we start with the following integral
\begin{equation}
\int_{S} \alpha^V_{ijklmn} c_{ij} c_l c_m c_n \textbf n_k dS
\end{equation}
Using Gauss theorem, we have 
\begin{eqnarray*}
\int_{S} \alpha^V_{ijklmn} c_{ij} c_l c_m c_n \textbf n_k dS &=& 
\int_{V} \frac{\partial \left[\alpha^V_{ijklmn}c_{ij} c_l c_m c_n \right]}{\partial x_k }  dV
\end{eqnarray*}
\begin{widetext}
\begin{equation}
\int_{S} \alpha^V_{ijklmn} c_{ij} c_l c_m c_n \textbf n_k dS = 
\int_{V} \frac{\partial{\alpha^V_{ijklmn}}}{\partial c} c_{ij} c_l c_m c_n c_k dV
+3 \int_{V} \alpha^V_{ijklmn} c_{ij} c_{kl} c_m c_n dV +\int_{V} \alpha^V_{ijklmn} c_{ijk} c_l c_m c_n dV
\end{equation}
\end{widetext}
Assuming the surface term to be zero, the above integral reduces to
\begin{widetext}
\begin{equation*}
\int_{V} \alpha^V_{ijklmn} c_{ijk} c_{l} c_m c_n dV=
-\int_{V} \frac{\partial{\alpha^V_{ijklmn}}}{\partial c} c_{ij} c_l c_m c_n c_k dV
-3\int_{V} \alpha^V_{ijklmn} c_{ij} c_{kl} c_m c_n dV\\
\end{equation*}
\end{widetext}

Thus, it is possible to drop the $\alpha^V_{ijklmn}$ term from the expansion and 
replace it with the two terms on the RHS of the equation above. 
Because of the intrinsic symmetry considerations (described in detail below), the terms on the RHS add to 
$\alpha^{II}_{ijklmn}$ and $\alpha^{III}_{ijklmn}$ terms respectively to give
\begin{equation*} 
\alpha^{III}_{ijklmn} = \alpha^{III}_{ijklmn}-3\alpha^V_{ijklmn}
\end{equation*}
and 
\begin{equation*} 
\alpha^{II}_{ijklmn} = \alpha^{II}_{ijklmn}-\frac{\partial}{\partial c}\alpha^V_{ijklmn}.
\end{equation*}

Note that for the sake of notational simplicity, we indicate the modified tensors also
using the same roman superscript. As indicated earlier 
the $\kappa^{II}_{ij}$ and $\beta^{IV}_{ijkl}$ terms can also be dropped. 
Thus, one obtains the following free energy expression:
\begin{widetext}
\begin{eqnarray*} \label{ReducedFreeEnergyDensity}
f\left(c, c_i, c_{ij}, c_{ijk}\right)& = & f|_{0} + \kappa^I_{ij} c_i c_j 
+ \beta^{I}_{ijkl} c_i c_j c_k c_l  +\beta^{II}_{ijkl} c_{ij} c_k c_l + \beta^{III}_{ijkl} c_{ij} c_{kl} \nonumber   \\
&+& \alpha^{I}_{ijklmn} c_i c_j c_k c_l c_m c_n +\alpha^{II}_{ijklmn} c_{ij} c_k c_l c_m c_n 
+ \alpha^{III}_{ijklmn} c_{ij} c_{kl} c_m c_n  \\ \nonumber
& + & \alpha^{IV}_{ijklmn} c_{ij} c_{kl} c_{mn}+
+ \alpha^{VI}_{ijklmn} c_{ijk} c_{lm} c_n + \alpha^{VII}_{ijklmn} c_{ijk} c_{lmn} \\ \nonumber   
\end{eqnarray*}
\end{widetext}
The free energy density, thus, consists of one second rank, three fourth
rank and six sixth rank (property) tensors.

As we show below in Section~\ref{Section4}, using the demand that the contribution of each of these tensor 
terms is positive definite, it can be shown that the tensors $\beta_{ijkl}^{II}$, $\alpha_{ijklmn}^{II}$, $\alpha_{ijklmn}^{IV}$ and $\alpha_{ijklmn}^{VI}$ are identically zero. Thus, the free energy expansion 
reduces to
\begin{widetext}
\begin{eqnarray}
f\left(c, c_i, c_{ij}, c_{ijk}\right)& = & \left [f\right]_{0} +\kappa^I_{ij} c_i c_j + \beta^{I}_{ijkl} c_i c_j c_k c_l  + \beta^{III}_{ijkl} c_{ij} c_{kl} \nonumber   \\
&+& \alpha^{I}_{ijklmn} c_i c_j c_k c_l c_m c_n + \alpha^{III}_{ijklmn} c_{ij} c_{kl} c_m c_n 
+ \alpha^{VII}_{ijklmn} c_{ijk} c_{lmn} \\ \nonumber   
\end{eqnarray}
\end{widetext}

From the free energy expression above, it is clear that when the Tayor series expansion is truncated at the second rank tensor terms, it can represented (effectively) in terms of the gradients alone. On the other hand, when we truncate at fourth rank tensor terms, there are effectively two terms; one is the term involving only the gradients; the other one is the term involving only the curvatures. Thus, when we truncate at the sixth rank terms, we may expect that there are effectively three terms; one is the term involving only the gradients; the second is the one involving only the curvatures; the third if the one involving only the aberration terms. However, as we noted above (and,
as we show below), the terms involving only the curvatures can be shown to be zero due to the demand of positive definiteness. On the other hand, in the sixth rank tensor terms, there is also a term involving two curvatures and two gradients, namely, $\alpha_{ijklmn}^{III}$; for the rest of this paper, we neglect the term and assume it to be identically zero. This assumption is physically unjustified; however, we make it for the sake of algebraic simplicity. Further, dropping this term results in retention of sixth rank tensors involving only gradients and aberrations;
in other words, by dropping this term, the retained terms can be seen as logical continuation
of the works of Cahn and Hilliard~\citep{Cahn-Hilliard} (who retained only gradient terms) and Abinandanan and Haider~\citep{AbiHaider} (who retained gradients and curvatures). Thus, the final free energy expression that we will 
use for the rest of this paper is as follows:
\begin{widetext}
\begin{eqnarray} \label{FinalFreeEnergy}
f\left(c, c_i, c_{ij}, c_{ijk} \right)& = & \left [f\right]_{0} +  \kappa^I_{ij} c_i c_j + \beta^{I}_{ijkl} c_i c_j c_k c_l  + \beta^{III}_{ijkl} c_{ij} c_{kl} \nonumber   \\ 
&+&\alpha^{I}_{ijklmn} c_i c_j c_k c_l c_m c_n + \alpha^{VII}_{ijklmn} c_{ijk} c_{lmn} \\ \nonumber   
\end{eqnarray}
\end{widetext}

\section{Symmetry considerations}\label{Section3}

The total number of components in a tensor of rank $n$ in a $d$ dimensional space is $d^n$. 
Thus, in 3 dimensions, the total number of components in the second, fourth and sixth rank tensors
are 9, 81, and 729, respectively. However, using symmetry considerations, the total number of 
non-zero components can be shown to be a much smaller number. Further, we can also show that,
of the non-zero components, only a few are independent. The symmetry arguments used to deduce 
the total number of non-zero and independent components are of two types, namely,
arguments based on intrinsic symmetry of the tensor itself and arguments based on the 
underlying crystalline symmetry -- as discussed in the following two subsections.

\subsection{Intrinsic symmetry arguments}

Consider the second rank tensor $\kappa^{I}_{ij}$; it multiplies $c_i c_j$. Since multiplication
is commutative, $c_j c_i$ is also the same as $c_i c_j$. In the free energy expansion, thus,
one can see that the terms $\kappa^{I}_{ij}$ and $\kappa^{I}_{ji}$ will always appear in the 
following combination, namely, $\kappa^{I}_{ij} + \kappa^{I}_{ji}$. Hence, without loss of generality, 
one can assume $\kappa^{I}_{ij}$ to be symmetric. 

Even though we have subsumed $\kappa^{II}_{ij}$ into $\kappa^{I}_{ij}$, it is possible
to argue that $\kappa^{II}_{ij}$ is symmetric using slightly different arguments; since it multiplies $c_{ij}$, and since for coarse-grained
composition fields and their higher order derivatives are continuous, $c_{ij}$ is the same as $c_{ji}$; hence, again, without loss of generality, one can assume that $\kappa^{II}_{ij}$ is symmetric. 

In other words, by invoking such, so-called intrinsic symmetries, we are able to reduce the total number of independent components of the second rank tensors from nine to six. 

Similar arguments can be used to reduce the total number of independent components for the fourth and sixth rank
tensors. In Table.~\ref{Table1}, we list the reduction in number of arguments of the different
tensors purely based on intrinsic symmetry considerations.

\begin{table}
\caption{\label{Table1} A table listing the number of independent components of the different tensor terms based on intrinsic symmetry considerations -- that is, indices that are interchangeable.}
\begin{tabular}{|l|c|c|c|}
\hline \hline
& & & \\
{\bf S. No.} & {\bf Tensor} & {\bf Intrinsic symmetry} & {\bf Number} \\
& & & {\bf of independent} \\
& & &  {\bf components}\\
& & & \\ 
\hline \hline
& & & \\
1 & $\kappa^{I}_{ij}$ & $i$ and $j$  & 6 \\
&&&\\
2 & $\beta^{I}_{ijkl}$ & All of $i$,$j$,$k$ and $l$   & 15\\ 
&&&\\
3 & $\beta^{III}_{ijkl}$ &$i$ and $j$, $k$ and $l$, & \\
& & and $ij$ and $kl$  & 21\\ 
&&&\\
4 & $\alpha^{I}_{ijklmn}$ & All the indices & 28 \\ 
&&& \\
5 & $\alpha^{VII}_{ijklmn}$ & $i$, $j$, and $k$, $l$, &\\
& & $m$, and $n$, & \\
& & and, $ijk$ and $lmn$ & 55\\
& & & \\
\hline \hline
\end{tabular}
\end{table}

\subsection{Crystalline symmetry arguments}

After reducing the number of independent components using intrinsic symmetry arguments, one can
reduce their number still further by considering the crystalline symmetry of the underlying continuum;
for example, if we consider a cubic symmetry for the underlying crystalline lattice, since
all second rank tensors are isotropic in a cubic crystal, one can see that there is only
one independent component (and three non-zero components) for the second rank tensors. The number 
of non-zero and independent components for isotropic, cubic and hexagonal 
symmetries for second and fourth rank tensors (of albeit only certain intrinsic symmetry) are
very well known and are listed in the classic textbook of Nye~\cite{Nye} for example. In this paper,
for the sake of completion we list them in Table.~\ref{Table2a} (second rank tensors) and 
Table.~\ref{Table2b} (fourth rank tensors).

\begin{table}
\caption{ \label{Table2a} A table listing the number of non-zero and independent components of second rank tensors
with different underlying crystalline symmetries. The matrices are symmetric; hence, only the diagonal terms and terms to the right of the diagonal are mentioned; components not-mentioned 
in the table are identically zero.}
\begin{tabular}{|l|c|c|}
\hline \hline
& &  \\
{\bf S. No.} & {\bf Crystalline symmetry} & {\bf The non-zero } \\
& & {\bf and independent} \\
& &   {\bf components}\\
& &  \\ 
\hline \hline
& &  \\
1 & Isotropic & $\kappa_{11}=\kappa_{22}=\kappa_{33}$\\
&&\\
2 & Cubic &  $\kappa_{11}=\kappa_{22}=\kappa_{33}$ \\ 
&&\\
3 & Hexagonal &  $\kappa_{11}=\kappa_{22}$\\ 
&& $\kappa_{33}$\\
\hline \hline
\end{tabular}
\end{table}

\begin{table}
\caption{ \label{Table2b} A table listing the number of non-zero and independent components of fourth rank tensors
with different underlying crystalline symmetries. The matrices are symmetric; hence, only the diagonal terms and terms to the right of the diagonal are mentioned; components not-mentioned 
in the table are identically zero.}
\begin{tabular}{|l|c|c|}
\hline \hline
& &  \\
{\bf S. No.} & {\bf Crystalline} & {\bf The non-zero and} \\
& {\bf Symmetry} & {\bf independent} \\
& &   {\bf components}\\
& &  \\ 
\hline \hline
& &  \\
1 & Isotropic &   $\beta^{I}_{11} = \beta^{I}_{22} = \beta^{I}_{33}$ \\
&&$\beta^{I}_{12} = \beta^{I}_{23} = \beta^{I}_{13} [= \beta^{I}_{11}]$ \\
&&\\
2 & Isotropic & $\beta^{III}_{11} = \beta^{III}_{22} = \beta^{III}_{33}$ \\
&&$\beta^{III}_{12} = \beta^{III}_{23} = \beta^{III}_{13}$ \\
&&$\beta^{III}_{44} = \beta^{III}_{55} = \beta^{III}_{66}  [= 2(\beta^{III}_{11} -\beta^{III}_{12} )]$\\
&&\\
3 & Cubic & $\beta^{I}_{11} = \beta^{I}_{22} = \beta^{I}_{33}$ \\
&&$\beta^{I}_{12} = \beta^{I}_{23} = \beta^{I}_{13}$ \\
&&\\
4 & Cubic & $\beta^{III}_{11} = \beta^{III}_{22} = \beta^{III}_{33}$ \\
&&$\beta^{III}_{12} = \beta^{III}_{23} = \beta^{III}_{13}$ \\
&&$\beta^{III}_{44} = \beta^{III}_{55} = \beta^{III}_{66}$\\
&&\\
4 & Hexagonal & $\beta^{I}_{11} = \beta^{I}_{22}$ \\
& & $\beta^{I}_{33}$ \\
&&$\beta^{I}_{12} [= \beta^{I}_{11}] $  \\
& &  $\beta^{I}_{23} = \beta^{I}_{13}$ \\
&&\\
4 & Hexagonal & $\beta^{III}_{11} = \beta^{III}_{22}$ \\
 & & $\beta^{III}_{33}$ \\
&&$\beta^{III}_{12}$ \\
  &&$\beta^{III}_{23} = \beta^{III}_{13}$ \\
&&$\beta^{III}_{44} = \beta^{III}_{55}$\\
&& $\beta^{III}_{66} [=2(\beta^{III}_{11}-\beta^{III}_{12})]$\\
&&\\
&&\\
\hline \hline
\end{tabular}
\end{table}

At this point, we wish to note that the second and fourth rank tensors are sometimes
represented by matrices. The matrix representation is fairly straight-forward in the case of 
second rank tensors. However, in the case of fourth rank tensors with intrinsic symmetry 
the same as $\beta^{III}$, the following transformations are used to reduce the fourth rank 
tensor with 81 terms to a matrix with $6 \times 6$ terms:
$11 \Rightarrow 1$; $22 \Rightarrow 2$; $33 \Rightarrow 3$; $23 \Rightarrow 4$; $13 \Rightarrow 5$;
and $12 \Rightarrow 6$ (See Nye~\cite{Nye} for details). 

Note that for $\beta^{I}$, the matrix representation is again in terms of a $3 \times 3$ matrix. To see this, consider the term in the free energy containing $\beta^{I}$. Its contribution to the expansion can be written in a matrix form as follows:

\hspace{50pt}
\begin{tabular}{c c c}
&&\\
  $ \left[ \begin{array}{ccc} 
 c^2_{1}  & c^2_{2}  &  c^2_{3} \end{array} \right] $ 
  & $ \left[ \begin{array}{ccc} 
\beta_{11}&  \beta_{12}&  \beta_{13} \\
\beta_{12}&           \beta_{22}&         \beta_{23} \\
\beta_{13}&           \beta_{23}&           \beta_{33}  \end{array} \right] $  & $ \left[ \begin{array}{c} 
 c^2_{1} \\
 c^2_{2} \\    
 c^2_{3} \end{array} \right]  $ \\ 
&&\\
\end{tabular}\\

This is because, in this tensor, components in which the indices occur odd number of times are identically zero for all the three cases that are considered here, namely, isotropic, cubic, and hexagonal. That is, the contribution to expansion from $\beta^I$ terms is the following:
\begin{eqnarray*}
&&\beta^{I}_{1111} c^4_1 + \beta^{I}_{2222}c^4_2  + \beta^{I}_{3333} c^4_3 \\
&&(\beta^{I}_{1122}+\beta^{I}_{2211}+ \beta^{I}_{1212} + \beta^{I}_{2121} + \beta^{I}_{1221} + \beta^{I}_{2112})c^2_1 c^2_2 \\
&&(\beta^{I}_{3311}+\beta^{I}_{1133}+ \beta^{I}_{1313} + \beta^{I}_{3131} + \beta^{I}_{1331} + \beta^{I}_{3113})c^2_1 c^2_3 \\
&&(\beta^{I}_{3322}+\beta^{I}_{2233}+ \beta^{I}_{2323} + \beta^{I}_{3232} + \beta^{I}_{2332} + \beta^{I}_{3223})c^2_2 c^2_3 
\end{eqnarray*}

where $\beta_{12} = \beta_{21}$ in the reduced matrix representation is $\frac{1}{2} \left[ \beta^{I}_{1122}+\beta^{I}_{2211}+ \beta^{I}_{1212} + \beta^{I}_{2121} + \beta^{I}_{1221} + \beta^{I}_{2112} \right]$, and so on. As noted in the intrinsic symmetry section above, further, we can assume all the six terms in the preceding expression to be identical without loss of generality. We also note that all our reduced representations are symmetric matrices.

\subsection{Sixth rank tensors and crystalline symmetry}

In this subsection, we explicitly show the deduction of the total number of independent
and non-zero components for the sixth rank terms assuming the crystalline symmetries of
hexagonal, cubic and isotropic and summarise the results in 
Tables.~\ref{Table2ca},~\ref{Table2cb}, and~\ref{Table2cc}.

In general, any point group symmetry can be characterized by a group of 
orthogonal transformations (represented by the matrices $a_{ij}$). 
Since by definition, tensors are quantities which
transform in a particular fashion under coordinate transformations, for a tensor to possess
the point group symmetry, in terms of the group of orthogonal
transformations that represent the point group symmetry, 
the following conditions are to be satisfied (for every $a_{ij}$ of the group) (See Nye~\cite{Nye}, for example):
\begin{equation} \label{TransformationEquation}
T_{ijk......}= a_{i\alpha} a_{j\beta} a_{k\gamma} ......T_{\alpha \beta \gamma .......}
\end{equation}
Using these conditions, we can identify the non-zero components, as well as the relationships between them, 
if any. The relationships, when identified, reduce the number of independent components.

With 729 components for the sixth rank tensor, the algebra of such a reduction is both formidable and 
laborious. However, in the literature on invariant theory (see for example Smith et al~\cite{Smith1962}),
the task has been made easier by listing of what is known as integrity bases. Using the integrity bases
(for example, using the Equation 5.17 in Smith et al~\cite{Smith1962}), we can write down the contribution of the sixth rank tensor to the free energy assuming hexagonal symmetry (albeit only for gradients and curvatures). For example, from Equation 5.17(ii) in Smith et al~\cite{Smith1962}, we see that the contribution from $\alpha^{I}_{ijklmn}$ term is a linear combination of $(c_3^2)^3$, $(c_1^2 + c_2^2)^3$, $c_1^2(c_1^2-3c_2^2)^2$, $(c_3^2)^2(c_1^2+c_2^2)$ and $c_3^2(c_1^2+c_2^2)^2$; the constant terms in the linear combination indicate the non-zero components and the relationship between the non-zero components. For example, from these terms, it is clear that terms of the type $\alpha_{111222}$ are identically zero (that is, in general, components with an index occurring odd number of times are zero); also, for example, with a little bit of algebra, one can show that $3 \alpha_{111122} = 3 \alpha_{222222} - 2 \alpha_{111111}$ and so on.

In the case of aberrancy terms, there is no such ready reckoner available for us. Hence, the algebra has to be carried out methodically. For example, using the Table 1 of Smith et al~\cite{Smith1962}, and carrying out the calculations of Eq.~\ref{TransformationEquation},
one can deduce the general result that in the sixth rank tensor, for hexagonal symmetry, the tensor terms in which the indices appear odd number of times (that is, terms of the type $\alpha^{VII}_{iiiiij}$ with $j \neq i$, $\alpha^{VII}_{iiijjj}$ with $j \neq i$, $\alpha^{VII}_{iiijjk}$ with $i \neq j \neq k$ etc) are identically zero. Thus, the number of non-zero components reduce to 183 from 729. Of these 183, the fact that there are only 19 non-zero components can be deduced from intrinsic symmetry arguments. These 19 terms can be represented in a $10 \times 10$ matrix form using the following transformations: $111 \Rightarrow 1$; $222 \Rightarrow 2$; $333 \Rightarrow 3$; $112 \Rightarrow 4$; $113 \Rightarrow 5$; $221 \Rightarrow 6$; $223 \Rightarrow 7$; $331 \Rightarrow 8$; $332 \Rightarrow 9$; and $123 \Rightarrow 0$. We note that this matrix representation is useful in the next section wherein constraints on the independent components of the tensors are derived. The total number of independent components in $\alpha^{VII}$ are 9. The reduction from 19 to 9 is obtained again using Table 1 of Smith et al~\cite{Smith1962} and Eq.~\ref{TransformationEquation}.

Similar operations can also be carried out to identify the number of independent and non-zero components for cubic symmetry. Once we have the non-zero and independent components listed for cubic and hexagonal symmetries, by looking at the intersection of these lists, the number of independent and non-zero components for the isotropic case for these sixth rank tensors can be identified. In Tables~\ref{Table2ca},~\ref{Table2cb},~\ref{Table2cc}, these results are 
summarised. 

\begin{table}
\caption{ \label{Table2ca} A table listing the number of non-zero and independent components of sixth rank tensors
for crystalline systems that are isotropic. The matrices are symmetric; hence, only the diagonal terms and terms to the right of the diagonal are mentioned; components not-mentioned 
in the table are identically zero.}
\begin{tabular}{|l|l|l|}
\hline \hline
& &  \\
{\bf S. No.} & {\bf Tensor} & {\bf The non-zero and independent} \\
& &   {\bf components}\\
& &  \\ 
\hline \hline
& &  \\
1 & $\alpha^{I}_{ijklmn}$ & $\alpha_{11} = \alpha_{22} = \alpha_{33}$ \\
& & $\alpha_{44} = \alpha_{55} = \alpha_{66} =\alpha_{77} = \alpha_{88} = \alpha_{99}[=\frac{9}{5} \alpha_{11}]$\\
& & $\alpha_{16} = \alpha_{18} = \alpha_{24} =\alpha_{29} = \alpha_{35} = \alpha_{37}$\\
& & = $\alpha_{49} = \alpha_{57} = \alpha_{68} [= \frac{3}{5} \alpha_{11}] $\\
& & $\alpha_{00}[=\frac{12}{5}\alpha_{11}]$\\
&&\\
\hline \hline
& &  \\
2 & $\alpha^{VII}_{ijklmn}$ & $\alpha_{11} = \alpha_{22} = \alpha_{33}$ \\
& & $\alpha_{44} = \alpha_{55} = \alpha_{66} =\alpha_{77} = \alpha_{88} = \alpha_{99}$\\
& & $\alpha_{24} = \alpha_{16} = \alpha_{29} =\alpha_{37} = \alpha_{35} = \alpha_{18}$ \\
& & $[= \frac{1}{2} (3 \alpha_{11} - \alpha_{44})]$ \\
& & $\alpha_{49} = \alpha_{68} = \alpha_{57} [= 2 \alpha_{18}]$ \\
& & $\alpha_{00} [= 2 (\alpha_{55} - 2 \alpha_{57})]$\\
&&\\
\hline \hline
\end{tabular}
\end{table}

\begin{table}
\caption{\label{Table2cb} A table listing the number of non-zero and independent components of sixth rank tensors
for crystalline systems that are cubic. The matrices are symmetric; hence, only the diagonal terms and terms to the right of the diagonal are mentioned; components not-mentioned 
in the table are identically zero.} 
\begin{tabular}{|l | c | c |}
\hline \hline
& &  \\
{\bf S. No.} & {\bf Tensor} & {\bf The non-zero and independent} \\
& &   {\bf components}\\
& &  \\ 
\hline \hline
& &  \\
1 & $\alpha^{I}_{ijklmn}$ & $\alpha_{11} = \alpha_{22} = \alpha_{33}$ \\
& & $\alpha_{44} = \alpha_{55} = \alpha_{66} =\alpha_{77} = \alpha_{88} = \alpha_{99}$\\
& & $\alpha_{00}$ \\
& & $\alpha_{16} = \alpha_{18} = \alpha_{24} =\alpha_{29} = \alpha_{35} = \alpha_{37}[=\frac{1}{3}\alpha_{44}]$\\
& & $\alpha_{49} = \alpha_{57} = \alpha_{68}[=\frac{1}{4}\alpha_{00}]$\\
&&\\
\hline \hline
& &  \\
2 & $\alpha^{VII}_{ijklmn}$ & $\alpha_{11} = \alpha_{22} = \alpha_{33}$ \\
& & $\alpha_{24} = \alpha_{16} = \alpha_{29} =\alpha_{37} = \alpha_{35} = \alpha_{18}$\\
& & $\alpha_{44} = \alpha_{55} = \alpha_{66} =\alpha_{77} = \alpha_{88} = \alpha_{99}$\\
& & $\alpha_{49} = \alpha_{68} = \alpha_{57}$ \\
& & $\alpha_{00}$ \\
&&\\
\hline \hline
\end{tabular}
\end{table}

\begin{table}
\caption{ \label{Table2cc} A table listing the number of non-zero and independent components of sixth rank tensors
for crystalline systems that are hexagonal. The matrices are symmetric; hence, only the diagonal terms and terms to the right of the diagonal are mentioned; components not-mentioned 
in the table are identically zero.}
\begin{tabular}{|l | c | c |}
\hline \hline
& &  \\
{\bf S. No.} & {\bf Tensor} & {\bf The non-zero and independent} \\
& &   {\bf components}\\
& &  \\ 
\hline \hline
& &  \\
1 & $\alpha^{I}_{ijklmn}$& $\alpha_{11}$ \\
&& $\alpha_{22}$ \\
&& $\alpha_{33}$ \\
&& $\alpha_{55} = \alpha_{77}$\\
&& $\alpha_{88} = \alpha_{99}$\\
&& $\alpha_{44}[=9\alpha_{22}-6\alpha_{11}] $ \\
&& $\alpha_{66}[=9\alpha_{11}-6\alpha_{22}] $ \\
&& $\alpha_{35} = \alpha_{37}[=\frac{1}{3}\alpha_{88}]$\\
&& $\alpha_{68} = \alpha_{57} = \alpha_{49} = \alpha_{29}= \alpha_{18}[=\frac{1}{3}\alpha_{55}]$ \\
&& $\alpha_{16} [ = \frac{1}{3} \alpha_{44}] $\\
&& $\alpha_{24} [ = \frac{1}{3} \alpha_{66}] $\\
&& $\alpha_{00} [ = \frac{4}{3} \alpha_{55}]$\\
&&\\
\hline \hline
& &  \\
2 & $\alpha^{VII}_{ijklmn}$& $\alpha_{11}$ \\
&& $\alpha_{33}$ \\
&& $\alpha_{44}$ \\
&& $\alpha_{66}$ \\
&& $\alpha_{57}$ \\
&& $\alpha_{18} = \alpha_{29}$ \\
&& $\alpha_{35} = \alpha_{37}$\\
&& $\alpha_{88} = \alpha_{99}$\\
&& $\alpha_{55} = \alpha_{77}$\\
&& $\alpha_{22} [ = \alpha_{11} + \frac{1}{9} (\alpha_{44} - \alpha_{66}) ] $ \\
&& $\alpha_{68} = \alpha_{49} [= \frac{1}{2} \alpha_{18}]$ \\
&& $\alpha_{16} [ = \frac{1}{2} (3 \alpha_{11} - \alpha_{44} ) ] $\\
&& $\alpha_{24} [ = \frac{1}{2} (3 \alpha_{11} - \frac{2}{3} \alpha_{66} - \frac{1}{3}\alpha_{44}) ] $\\
&& $\alpha_{00} [= 2 (\alpha_{55} - \alpha_{57})]$\\
&&\\
\hline \hline
\end{tabular}
\end{table}

\section{Constraints on the independent tensor terms}\label{Section4}

We have looked at the recipe to include higher order tensor terms in the free energy expansion.
These higher order terms contribute to the interfacial energy of the system as well as make 
it anisotropic. We can derive the constraints on these tensor components using the following argument. 
We demand the energy associated with an elemental volume with a non uniform concentration is greater 
than that with a uniform concentration (the uniform concentration being same as that of the average 
concentration of the elemental volume). This ensures that the contribution to the free energy from 
the interfaces is always positive; we also demand the positive definiteness of interfacial energy 
for each tensor term individually. These are stronger conditions; it might be possible to choose 
the different tensors and their independent components by considering all the tensor terms together 
to give positive definite energy for the interface. However, for the sake of simplicity and the freedom
in numerical implementation to include any of the terms independently, we demand 
elemental and term by term positive definiteness.

The demand of term by term positive definiteness leads to the conclusion that $\beta^{II}_{ijkl}$
is identically zero; to see this consider the term $\beta^{II}_{ijkl}c_{ij} c_k c_l$. Let us 
contract $\beta^{II}$ with the gradient terms and refer to the resultant second rank tensor 
as $\beta^{II}_{r}$; since $c_{ij}$ is arbitrary (and is 
independent of the gradients), the only way this term will give a positive definite contribution is by making the 
$\beta^{II}_{ijkl}$ term identically zero. Thus, the total number of 
fourth rank tensors reduce from three to two. 

Using similar arguments, it is also possible to show that the sixth rank tensors $\alpha^{II}_{ijklmn}$ and 
$\alpha^{VI}_{ijklmn}$ are identically zero. In addition, it is also possible to show that the demand of positive 
definiteness results in $\alpha^{IV}$ being identically zero; however, to do so, we need to use the invariant theory 
results from Smith et al~\cite{Smith1962}, generate the polynomial which results from the contribution of this term, and, 
exploit the arbitrariness in the choice of curvature terms to choose appropriate terms and hence show that each of the 
terms in the polynomial are identically zero; see the Appendix for the details of the algebra. 
Thus, the total number of sixth rank tensors reduced from six to three. 
Further, as noted above, for the sake of algebraic simplicity and logical continuity, we also assume that one of the 
sixth rank tensors, namely, $\alpha^{III}_{ijklmn}$ is identically zero. Thus, the total number of sixth rank tensors 
are finally reduced to two.

As noted in Nye~\cite{Nye}, the necessary and sufficient condition for positive definiteness of symmetric tensors is that 
in their matrix representation, the leading minors 
should be positive. Hence, using the matrix representations and demanding that the leading minors are positive, we can 
obtain the constraints on the independent components of the tensors. While this methodology works well for $\kappa^{I}$, 
$\beta^{III}$ and $\alpha^{VII}$, for the other
tensors in the free energy expansion, such a conditions is stronger. However, using the polynomial representations 
associated with these tensors, and demanding that they be always positive definite, the constraints can be derived; the 
derivations involve very simple algebraic manipulations and the notion of Lagrange multipliers for constrained 
optimization; see the Appendix for the details of the algebra. We have also used the results on positivity of cubic polynomials; specifically, Eq. 2.17 of Schmidt and 
Hess~\cite{SchmidtHess}.

In Tables~\ref{Table3a},~\ref{Table3b},~\ref{Table3ca},~\ref{Table3cb} and~\ref{Table3cc} the constraints on the independent components are listed. 

\begin{figure}
\includegraphics{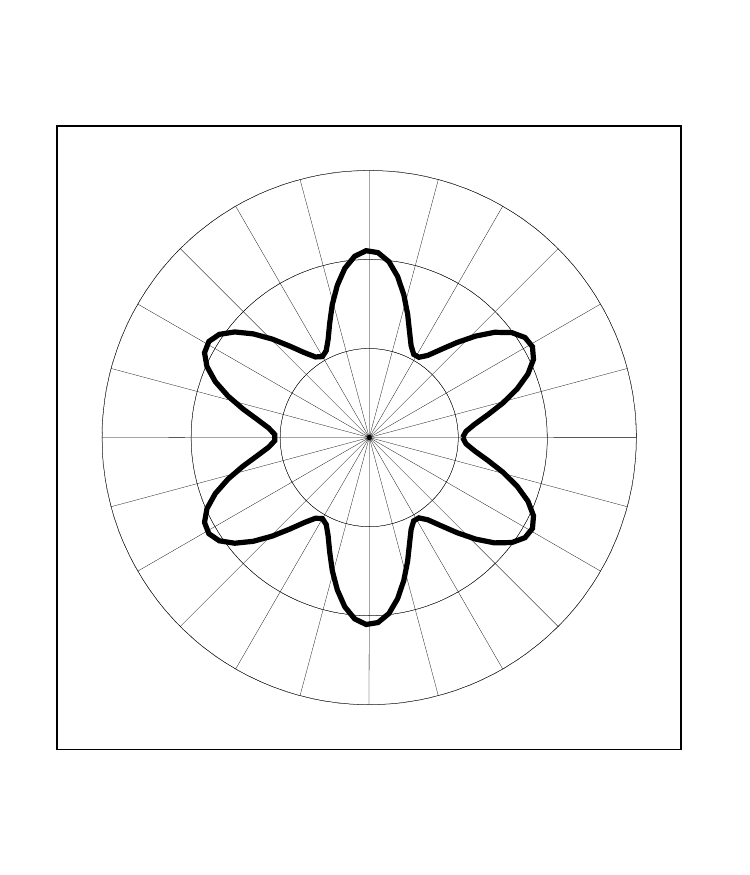}
\caption{\label{Figure1} Six-fold anisotropy of the basal plane obtained using the $\alpha^{I}$ term. When considering the hexagonal symmetry and only the basal plane, there are only two independent constants; they are chosen to be 2.1 and -1.05.}
\end{figure}

\begin{figure}
\includegraphics[scale=0.5]{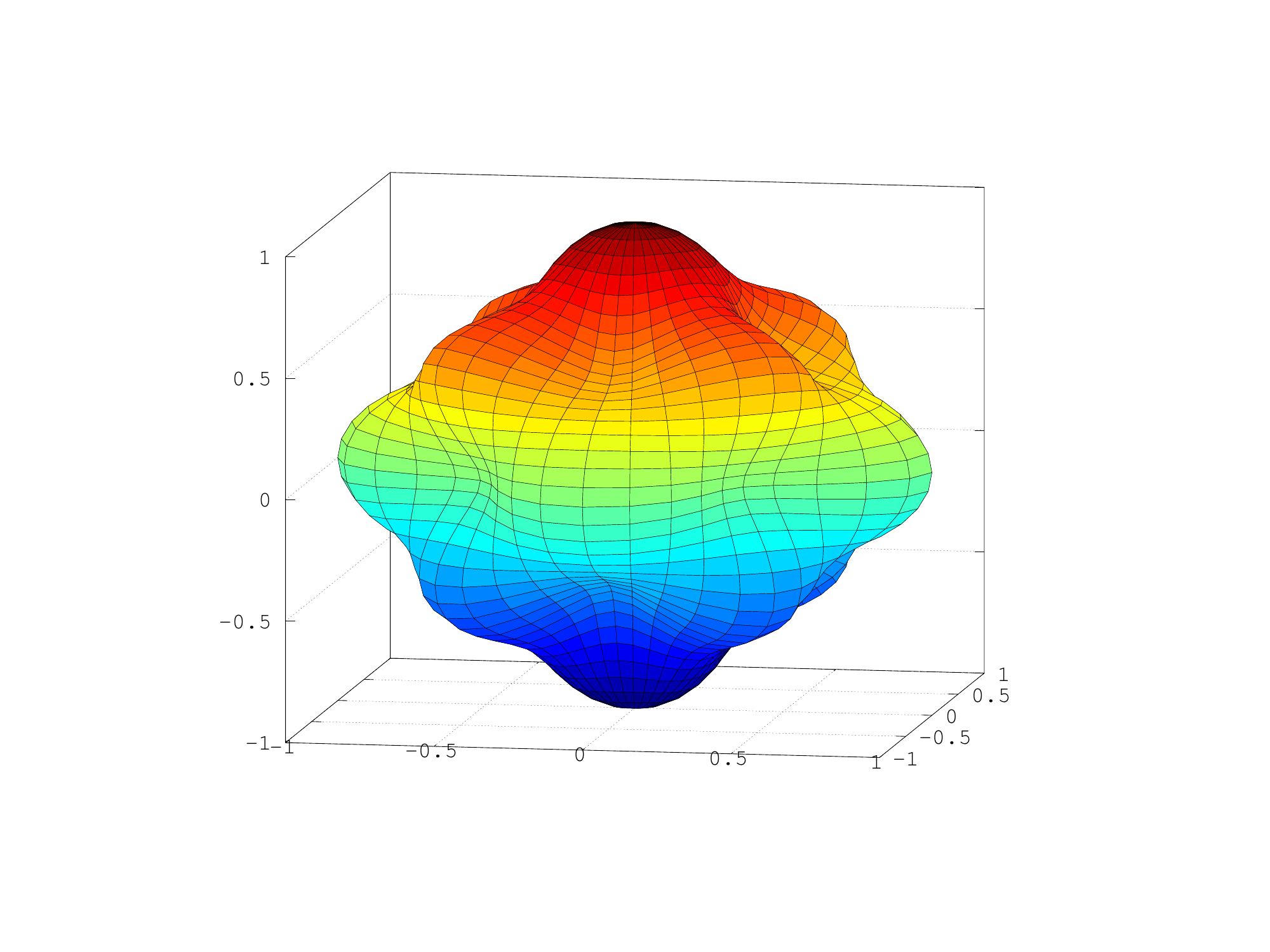}
\caption{\label{Figure2} Preference of $<110>$ directions over both $<100>$ and $<111>$ directions in a cubic system obtained using the $\alpha^{I}$ term. When considering cubic symmetry for the sixth rank tensor, there are only
three independent constants; they are chosen to be 1, -1.5 and 12.}
\end{figure}

With these constraints given, it is possible to choose the appropriate constants
and hence get the required anisotropy in the interfacial energy. For example, as noted in 
the introduction, the inclusion of the sixth rank tensor allows us to obtain hexagonal 
anisotropy and make cubic systems prefer $<110>$ over both $<100>$ and $<111>$. In Fig.~\ref{Figure1}, for example,
we show the $\alpha^{I}$ term plotted in the basal plane; in the basal plane, there are only two independent constants; for our plot, they are chosen to be 2.1 and -1.05. From the figure, it is clear that the six-fold anisotropy of the
basal plane is captured by the inclusion of this tensor term. If the parameters are chosen to be 2.1 and +1.05, the
six-fold anisotropy is retained with the directions of minimum energy rotated in the plane by 30$^{\circ}$.  In Fig.~\ref{Figure2}, we show the plot of $\alpha^{I}$ term with the independent constants chosen to be consistent with cubic anisotropy. Specifically, the three independent constants are chosen to be 1, -1,5 and 12. This choice of 
constants results in the preference of $<110>$ directions over both $<100>$ and $<111>$ directions as shown.
In plotting the above shapes, we have assumed that the interface profile is the same
in all directions; this is not strictly true; however, our preliminary numerical calculations have confirmed 
that in spite of the small changes in profiles of the interface along different directions, the overall 
anisotropy of the interfacial energy is consistent with what is shown in these figures.

\begin{table}
\caption{ \label{Table3a}A table listing the constraints on the independent components of second rank tensors
with different underlying crystalline symmetries.}
\begin{tabular}{|l | c | c |}
\hline \hline
& &  \\
{\bf S.} & {\bf Crystalline} & {\bf Constraints} \\
{\bf No.} & {\bf symmetry} &   \\
& &  \\ 
\hline \hline
& &  \\
1 & Isotropic & $\kappa_{11} > 0$\\
&&\\
2 & Cubic &  $\kappa_{11} > 0$ \\ 
&&\\
3 & Hexagonal &  $\kappa_{11} > 0$\\ 
&& $\kappa_{33} > 0$\\
\hline \hline
\end{tabular}
\end{table}

\begin{table}
\caption{ \label{Table3b} A table listing the constraints on the independent components of fourth rank tensors
with different underlying crystalline symmetries.}
\begin{tabular}{|l | c | c |}
\hline \hline
& &  \\
{\bf S.} & {\bf Crystalline} & {\bf Constraints} \\
{\bf No.} & {\bf symmetry} &   \\
& &  \\ 
\hline \hline
& &  \\
1 & Isotropic &   $\beta^{I}_{11} > 0 $ \\
&&\\
2 & Isotropic & $\beta^{I}_{11} > 0$ \\
&&$-\frac{\beta^{III}_{11}}{2} < \beta^{III}_{12} < \beta^{III}_{11}$ \\
&&\\
3 & Cubic & $\beta^{I}_{11} > 0$ \\
&& $\beta^{I}_{12} > - \frac{\beta^{I}_{11}}{2}$ \\
&&\\
4 & Cubic & $\beta^{I}_{11} > 0$ \\
&&$-\frac{\beta^{III}_{11}}{2} < \beta^{III}_{12} < \beta^{III}_{11}$ \\
&& $\beta^{III}_{44} > 0$\\
&&\\
5 & Hexagonal & $\beta^{I}_{11} > 0 $ \\
& & $\beta^{I}_{33} > 0$ \\
&& if $\beta^{I}_{13} < 0$, then \\
&& $(\beta^{I}_{13})^2 - \beta^{I}_{11} \beta^{I}_{33} < 0 $, and \\
&& $\beta^{I}_{13} > -\frac{1}{2} \left[ \beta^{I}_{11} \left( \frac{\beta^{I}_{33} - \beta^{I}_{13}}{\beta^{I}_{11} - \beta^{I}_{13}} \right) + \beta^{I}_{33} \left( \frac{\beta^{I}_{11} - \beta^{I}_{13}}{\beta^{I}_{33} - \beta^{I}_{13}} \right) \right]$ \\
&&\\
6 & Hexagonal & $\beta^{III}_{11} > 0$ \\
&& $-\beta^{III}_{11} < \beta^{III}_{12} < \beta^{III}_{11}$ \\
&&$\beta^{III}_{44} > 0$ \\
&& $(\beta^{III}_{11} +\beta^{III}_{12})\beta^{III}_{33} > 2 (\beta^{III}_{13})^2$\\
&&\\
&&\\
\hline \hline
\end{tabular}
\end{table}

\begin{table}
\caption{ \label{Table3ca}A table listing the constraints on the independent components of sixth rank tensors
assuming isotropy.}
\begin{tabular}{|l | c | c |}
\hline \hline
& &  \\
{\bf S.} & {\bf Tensor} & {\bf Constraints} \\
{\bf No.} & & \\
& &  \\ 
\hline \hline
& &  \\
1 & $\alpha^{I}_{ijklmn}$ & $\alpha_{11} >0$ \\
&&\\
\hline \hline
& &  \\
2 & $\alpha^{VII}_{ijklmn}$ & $\alpha_{11} > 0 $ \\
& & $\left(\frac{15-\sqrt(105)}{4}\right)\alpha_{11} < \alpha_{44} < \left(\frac{15+\sqrt(105)}{4}\right)\alpha_{11}$\\
&&\\
\hline \hline
\end{tabular}
\end{table}

\begin{table}
\caption{ \label{Table3cb}A table listing the constraints on the independent components of sixth rank tensors
assuming cubic anisotropy.}
\begin{tabular}{|l | c | c |}
\hline \hline
& &  \\
{\bf S.} & {\bf Tensor} & {\bf Constraints} \\
{\bf No.}& & \\
& &  \\ 
\hline \hline
& &  \\
1 & $\alpha^{I}_{ijklmn}$ & $\alpha_{11} > 0 $ \\
& & $\frac{5}{3} \alpha_{44} > -\alpha_{11}$\\
& & $\alpha_{00} > - 6 \alpha_{44}$ \\
&&\\
\hline \hline
& &  \\
2 & $\alpha^{VII}_{ijklmn}$ & $\alpha_{11} > 0$ \\
& & $\alpha_{11}  \alpha_{44} > (\alpha_{24})^2$\\
& & $(\alpha_{44} - \alpha_{57}) \Lambda > 0$\\
& & where $\Lambda = (\alpha_{11} \alpha_{44} - 2(\alpha_{24})^2 +  \alpha_{11}\alpha_{57})$\\
& & $\alpha_{00}>0$ \\
&&\\
\hline \hline
\end{tabular}
\end{table}

\begin{table}
\caption{ \label{Table3cc}A table listing the constraints on the independent components of sixth rank tensors
assuming hexagonal anisotropy. We have assumed that $\alpha^{I}_{22} < \alpha^{I}_{11}$; if $\alpha^{I}_{22} > \alpha^{I}_{11}$, in this expression, $\alpha^{I}_{22}$ will be replaced by $\alpha^{I}_{11}$. Also see~\cite{SchmidtHess}.}
\begin{tabular}{|l | c | c |}
\hline \hline
& &  \\
{\bf S.} & {\bf Tensor} & {\bf Constraints} \\
{\bf No.} & & \\
& &  \\ 
\hline \hline
& &  \\
1 & $\alpha^{I}_{ijklmn}$& $\alpha_{11} > 0$ \\
&& $\alpha_{22} > 0$ \\
&& $\alpha_{33} > 0$ \\
&& $\alpha_{55} > 0$ and $\alpha_{88} > 0$ or,\\
&& $\frac{500}{9} \alpha_{22} (\alpha_{88})^3 + \frac{500}{9}\alpha_{33} (\alpha_{55})^3$ \\
&& $+ 27 (\alpha_{22})^2 (\alpha_{33})^2- 50 \alpha_{22} \alpha_{33} \alpha_{55}$ \\ 
&& $\alpha_{88} - \frac{625}{81}(\alpha_{55})^2 (\alpha_{88})^2 > 0$ \\
&&(See also the caption).\\
&&\\
\hline \hline
& &  \\
2 & $\alpha^{VII}_{ijklmn}$& $\alpha_{11} > 0$ \\
&& $\alpha_{22} > 0$ \\
&& $\alpha_{33} > 0$ \\
&& $\alpha_{22} \alpha_{44} > (\alpha_{24})^2$ \\
&& $\alpha_{33} \alpha_{55} > (\alpha_{35})^2$ \\
&& $\alpha_{11} \alpha_{66} > (\alpha_{16})^2$ \\
&& $\Lambda_1 > 0$ \\
&& where $\Lambda_1 = \alpha_{22} \alpha_{44} \alpha_{88}$ \\
&& $- \alpha_{22} (\alpha_{68})^2-\alpha_{44} (\alpha_{18})^2$\\
&& $+ 2 \alpha_{68} \alpha_{18} \alpha_{24}-\alpha_{88} (\alpha_{24})^2$ \\
&& $\Lambda_2 > 0$ \\
&& where $\Lambda_2 = \alpha_{11} \alpha_{66} \alpha_{88}$ \\
&& $- \alpha_{11} (\alpha_{68})^2-\alpha_{66} (\alpha_{18})^2$\\
&& $+ 2 \alpha_{68} \alpha_{18} \alpha_{16}-\alpha_{88} (\alpha_{16})^2$ \\
&& $\Lambda_3 > 0$ \\
&& where $\Lambda_3 = \alpha_{33} [(\alpha_{55})^2 - (\alpha_{57})^2]$ \\
&& $+ 2 \alpha_{57} \alpha_{37} \alpha_{35}$ \\
&& $- \alpha_{55} [(\alpha_{37})^2 + (\alpha_{35})^2]$ \\
&& $\alpha_{00} >0$ \\
&&\\
\hline \hline
\end{tabular}
\end{table}

\section{Invariant theory and higher order tensors}\label{Section5}

The algebra associated with sixth rank tensor terms is very laborius and cumbersome as it involves 729
components in three dimensions. However, it is possible to deal with the sixth rank tensor terms 
without carrying out any explicit calculations. For doing so, we expand the tensor terms and compare 
them to the corresponding polynomials constructed out of the integrity basis (which are well documented 
in the literature of theory of invariants for any given point group symmetry) (as long as only gradient and curvature terms are considered). For example, in the following sub-section, we show how the integrity basis lists 
help us in constructing the free energy expansion in a tensor term that involves only gradients.
In the case of tensors that involve aberrancy terms, however, the following methodology cannot be used: the explicit 
tensor algebra has to be carried  out to construct the polynomials.

\subsection{Extending free energy expansion to higher order tensor terms using integrity basis} \label{SixthRankTensorPolynomial}

Consider the term $\alpha^I_{ijklmn} c_i c_j c_k c_l c_m c_n$ in the free energy expansion.
This is a polynomial in the gradients of concentration. The coefficients 
in this multivariate polynomial are linear combinations of components of the tensor $\alpha^I$. 
Thus if one knows this polynomial, one can deduce the zero and independent components of the tensor without
doing the laborious tensor decomposition of Eq.~\ref{TransformationEquation}. The integrity bases help us 
in constructing precisely these polynomials. An integrity basis is a list of 
polynomials which are invariant under a particular group of transformations 
and every other polynomial invariant under this group of transformations can be built out of 
them by the operations of multiplication and addition among them.

Let us consider $\alpha^I_{ijklmn}$ for the case of Hexoctahedral symmetry. Its components
multiply the gradient terms in the expansion to form a sixth order polynomial. This polynomial 
can be constructed from the integrity basis of Hexoctahedral class listed in page number 18 of 
Smith et al~\cite{Smith1962}. We see from the list that only $I_{10}$ , $I_{11}$, $I_{12}$
contain only vector terms. We can form a polynomial
of degree six out of these building blocks as follows: 
$(I_{10})^3$ , $I_{10}I_{11}$, $I_{12}$. Thus a sixth order polynomial in the 
components of the gradient of composition which is invariant under the orthogonal transformations 
of Hexoctahedral group is: $ p I_{10}^3 +  q I_{10}I_{11} + rI_{12}$, where $p$, $q$ and $r$ are 
constants; since $I_{10}= (c_1^2 + c_2^2 +c_3^2)^3$, 
$I_{11} = (c^2_1c^2_2+c^2_2c^2_3+c^2_3c^2_1)$ and $i_{12} = (c^2_1c^2_2c^2_3)$, 
we see that the sixth order polynomial consists of the following three terms (which are
multiplied by $p$, $3p+q$ and $6p+3q+r$, respectively):\\ 
$c^6_1+c^6_2+c^6_3$\\
$c^4_1c^2_2+c^4_1c^2_3+c^4_2c^2_1+c^4_2c^2_3+c^4_3c^2_1+c^4_3c^2_2$\\
$c^2_1c^2_2c^2_3$.

On the other hand $\alpha^I_{ijklmn} c_i c_j c_k c_l c_m c_n$ is also a sixth order polynomial that
should also be invariant under these very transformations. In other words, the two are identical.
Hence, we obtain the non-zero components of the tensor $\alpha^I_{ijklmn}$ and the relations 
among them by equating the above expressions. 

Firstly, the coefficients of $c^6_1$, $c^6_2$ and $c^6_3$ 
are the same. But $\alpha^I_{111111}$, $\alpha^I_{222222}$ and $\alpha^I_{333333}$ multiply 
$c^6_1$, $c^6_2$ and $c^6_3$ respectively. Hence, these tensor components are equal in magnitude.

Secondly, we see from the above equation that a tensor component with an index occurring odd number 
of times is identically zero. This is because in the polynomial no component of gradient of 
composition occurs odd number of times.

Further, from the second term of the above polynomial we see that $\alpha^I_{111122}$=$\alpha^I_{111133}$
=$\alpha^I_{222211}$=$\alpha^I_{222233}$=$\alpha^I_{333311}$=$\alpha^I_{333322}$. Due to the internal symmetry of
this tensor all the components formed from the permutations of the indices of each of the term in the above equality 
are equal to those formed from the permutations of the indices of any of the term.

\section{Phase field evolution equation}\label{Section6}

Given the free energy (~\ref{FinalFreeEnergy}), namely,
\begin{eqnarray}
f\left(c, c_i, c_{ij}, c_{ijk}\right)& = & \left[ f \right]_{0} + \kappa^I_{ij} c_i c_j \nonumber \\
&+& \beta^{I}_{ijkl} c_i c_j c_k c_l  + \beta^{III}_{ijkl} c_{ij} c_{kl} \nonumber   \\
&+&\alpha^{I}_{ijklmn} c_i c_j c_k c_l c_m c_n \nonumber \\
&+&\alpha^{VII}_{ijklmn} c_{ijk} c_{lmn} 
\end{eqnarray}
we can obtain the evolution equation using the Euler-Lagrange equation of $F = \int f dV$; this equation is as follows:
\begin{equation} \label{EvolEquation}
\frac{\partial c}{\partial t}  = \nabla \cdot M \nabla \mu
\end{equation}
where $\mu$ is the chemical potential which is the variational derivative, $\delta F/\delta c$:
\begin{eqnarray} \label{ChemPot}
\mu = \frac{\delta F}{\delta c}& = & \frac{\partial \left[f\right]_{0}}{\partial c} \nonumber  \\
&-& 2 \kappa^I_{ij} c_{ij} \nonumber \\ 
&-&12 \beta^{I}_{ijkl} c_{ij} c_k c_l  + 2 \beta^{III}_{ijkl} c_{ijkl} \nonumber   \\
&-&30 \alpha^{I}_{ijklmn} c_{ij} c_k c_l c_m c_n - 2 \alpha^{VII}_{ijklmn} c_{ijklmn} \\ \nonumber   
\end{eqnarray}
Corresponding to the choice of the tensors (that is, isotropic, cubic or hexagonal), one can 
then obtain the corresponding interfacial energy anisotropy. The numerical implementation of the 
above equation using explicit and semi-implicit Fourier spectral technique in 1-, 2- and 3-D are in progress.

\section{Conclusions}

When the free energy is assumed to be a function {\bf only} of (coarse-grained) composition and its local gradient, curvature and aberration, there are seven sixth rank tensors in the Taylor series expansion. These seven tensors can be reduced to six using Gauss theorem. If we demand that the contribution of each of these tensor terms to the free energy is positive definite, the number of sixth rank tensors can be further reduced to three. Of these three tensors, we have decided to retain only the tensors that are associated with only the gradients and only the aberrancy terms. 
We have identified the total number of non-zero and independent components of these 
tensors by accounting for the intrinsic symmetries and the symmetry of the underlying continuum (isotropic, 
cubic and hexagonal); specifically, the number of independent components is very small (one or two isotropic, 
three or five for cubic, and, five or nine for hexagonal systems). In addition, we have also identified the 
constraints that these independent terms have to obey (under the condition that each 
of the tensor terms, when incorporated individually, always result in interfacial energies that are 
positive definite). Using the results from invariant group theory, we show that representation of these 
tensor terms in polynomial form is possible; in numerical implementations of the phase field model, such
polynomial forms can be quite handy. Finally, we show the phase
field evolution equations that follow from the free energy functional based on the given free energy density;
one of the sixth rank tensors leads to a linear term in the evolution equation while the other leads to a non-linear
term. Further work on numerical implementation as well as evaluating the tensor terms from other models and/or
experiments is in progress.

\begin{acknowledgments}
We thank T A Abinandanan and Arka Lahiri of Department of Materials Engineering, Indian Institute of Science, Bangalore for useful discussions and IRCC, IIT-Bombay for financial support through 09IRCC16.
\end{acknowledgments}

\appendix*

\section{Algebraic details of derivation of constraints}

\subsection{To show that $\alpha^{IV}_{ijklmn}$ is identically zero}

The components of the tensor $\alpha^{IV}_{ijklmn}$ multiply the curvature tensors, $c_{ij}$, $c_{kl}$, 
and $c_{mn}$; the resultant polynomial is a third degree polynomial in curvature terms. This polynomial
can be directly written from the integrity basis tables. 

Consider the case of dihexagonal dipyramidal class. The integrity 
basis list for this class is given on p. 106 of Smith et al~\cite{Smith1962} (with the building
blocks denoted by R). In the listing, we are not concerned with $R_{4}^2$, $R_{4}R_{8}$ and $R_{8}^2$ 
because they contain more than three curvature terms whereas 
we need to construct only  a polynomial of degree three.

Thus for the case of the symmetry class considered the polynomial 
$\alpha^{IV}_{ijklmn}c_{ij}c_{kl}c_{mn}$ is as follows:
\begin{eqnarray*}
&& A\;c^{2}_{33} +B\;(c_{11}+c_{22})^3 + C\;(c_{11}+c_{22})\;c^{2}_{33} \\ \nonumber
&&+D\;(c_{11}+c_{22})^2\;c_{33} + E\;(c_{11}c_{22}-c^2_{12})\;c_{33} \\ \nonumber
&& +F\;(c_{11}c_{22}-c^2_{12})(c_{11}+c_{22}) +G\;(c^2_{31}+c^2_{23})\;c_{33} \\ \nonumber
&& +H\;(c^2_{31}+c^2_{23})(c_{11}+c_{22}) +I\;c_{11}\;[(c_{11}+3c_{22})^2-12c^2_{12}] \\ \nonumber
&&+J\;(c_{22}c^2_{31}+c_{11}c^2_{23}-2c_{12}c_{23}c_{31}) 
\end{eqnarray*}

We want the above polynomial to be positive semi-definite. i.e., only when all the arguments $c_{ij}$ 
are zero do we want the polynomial to be zero and for all other possibilities we want it to be greater than 
or equal to zero. Choosing certain combinations of numerical values in a particular order as follows, 
we can show that all the coefficients in this polynomial are identically zero.

If $c_{33}$ is non zero and rest all are zeroes, then the above polynomial reduces to
$A\;c_{33}^3$. This can be greater than or equal to zero for all values of $c_{33}$ only when 
$A$ is zero. Hence, we are left with a polynomial which is same as above except now we don't have the first term.
Next, we consider the case when $(c_{11}+c_{22})$ is non zero and rest all are zeroes.
Then only the $I\;c_{11} [(c_{11}+3c_{22})^2-12c^2_{12}]$ term remains and it further reduces to
$I\;c_{11} (c_{11}+3c_{22})^2$ since $c_{12}$ is zero.
This further simplifies to 
$I\;c_{11} (c_{11}+c_{22}+2c_{22})^2$
since $(c_{11}+c_{22})=0$ we have
$I\;c_{11} (2\;c^2_{22})$
This can be greater than or equal to zero only when $I$ is identically zero. Thus we 
now have a polynomial which is same as what we started out with except there aren't the terms multiplying
$A$ and $I$.
Similarly we have the following results
$$c_{11}\neq0 \textrm{ and rest all are zereos} \implies B=0$$
$$c_{12}\neq0, c_{33}\neq0 \textrm{ and rest all are zereos} \implies E=0$$
$$c_{12}\neq0, c_{11}\neq0 \textrm{ and rest all are zereos} \implies F=0$$
$$c_{31}\neq0, c_{33}\neq0 \textrm{ and rest all are zereos} \implies G=0$$
$$c_{31}\neq0, c_{11}\neq0 \textrm{ and rest all are zereos} \implies H=0$$
$$c_{31}\neq0, c_{22}\neq0 \textrm{ and rest all are zereos} \implies J=0$$
Finally we are left with 
$C\;(c_{11}+c_{22})c^{2}_{33}+D\;(c_{11}+c_{22})^2c_{33}$
if $(c_{11}+c_{22})<0, c_{33}<0 \textrm{ and rest all are zereos}$ then we need $C\leq0 \textrm{ and } D\leq0$
but if instead $(c_{11}+c_{22})>0, c_{33}>0 \textrm{ and rest all are zereos}$ then we need $C\geq0 \textrm{ and } D\geq0$.
The above two are satisfied only if $C=0$ and $D=0$ identically. Thus the entire polynomial is identically zero.

\subsection{Derivation of the constraints on the tensor terms: details of the algebra for cubic symmetry of $\alpha^{I}_{ijklmn}$}

The components of the tensor $\alpha^{I}_{ijklmn}$ multiplies $c_{i}c_{j}c_{k}c_{l}c_{m}c_{n}$, 
the resultant function is a sixth degree polynomial in gradient terms. As noted in 
Sec.~\ref{SixthRankTensorPolynomial}, for Hexoctahedral systems, it is the following expression:
\begin{eqnarray*} \label{PDPolynomial}
I&=&A\;(c_1^2+c_2^2+c_3^2)^3 \\ \nonumber
&+& B\;(c_1^2+c_2^2+c_3^2)(c_1^2c_2^2+c_2^2c_3^2+c_3^2c_1^2)\\ \nonumber
&+& C\;c_1^2c_2^2c_3^2
\end{eqnarray*}

Our demand is that the above equation be positive semi-definite; that is, we demand $I=0$ only when 
all the $c_{i}$s are simultaneously zeroes and for all other possibilities $I\geq0$. This imposes 
certain constraints on $A$, $B$ and $C$. We obtain these constraints using the following procedure.

First, we consider a case in which only $c_1\neq0$ and all the remaining gradients are zero. For
Eq.~\ref{PDPolynomial} to be greater than or equal to zero for all values of $c_1 \ neq 0$,
we see that 
\begin{equation} \label{ConstrainForA}
  A \geq 0
\end{equation}

Now, we consider a case in which $c_1\neq0$ and $c_2\neq0$ and all the other gradients are
zero. This gives

$$I=A\;(c_1^2+c_2^2)^3+B\;(c_1^2+c_2^2)c_1^2c_2^2$$
$$\implies I=(c_1^2+c_2^2)^3\left(A\;+B\;\frac{c_1^2c_2^2}{(c_1^2+c_2^2)^2}\right) $$

We know that $\frac{c_1^2c_2^2}{(c_1^2+c_2^2)^2}$ is always less than or 
equal to $\frac{1}{4}$. Thus for the above equation to be greater than or equal to zero we need
\begin{equation} \label{ConstraintForB}
  B\geq -4A
\end{equation}

Finally, for the case of $c_1\neq0$, $c_2\neq0$ and $c_3\neq0$, we use the following approach:
find the global minimum of $I$ and demand that it be greater than or equal to zero. 
Finding the global minimum of the above multivariate polynomial $I$ is a little hard. Instead, 
we take a different approach which is easy and serves the purpose. We use the method of 
Lagrange multipliers for constrained extremization. We constrain $(c_1,c_2,c_3)$ to a sphere 
of radius $k$, we find the global minimum in this constrained region and demand it be greater than or equal to 
zero. Further we demand this for all the spheres. i.e., for all $k\in(0,\infty)$.

Let $G=c_1^2+c_2^2+c_3^2$. To find the values of $c_1$, $c_2$, $c_3$ which extremizes 
$I$ under the constraint $G=k^2$ we need to solve the following four equations in four 
unknowns (namely $c_1$, $c_2$, $c_3$ and $\lambda$).

$$c_1^2+c_2^2+c_3^2=k^2$$
$$\nabla I=\lambda \nabla G $$
The second of the above two equations is actually a set of three equations 
$\frac{\partial I}{\partial c_i} = \lambda \; \frac{\partial G}{\partial c_i} ; i=1,2,3$.\\

For $i=1$ we have 
\begin{eqnarray} \label{FirstEquation}
\lambda & = & 3A (c_1^2+c_2^2+c_3^2)^2 + B(c_1^2c_2^2+c_2^2c_3^2+c_3^2c_1^2) \\ \nonumber
& & +B(c_1^2+c_2^2+c_3^2)(c_2^2+c_3^2)+ C\;c_2^2c_3^2 
\end{eqnarray}

Similarly for $i=2$ and $i=3$ we have
\begin{eqnarray} \label{SecondEquation}
\lambda & = & 3A (c_1^2+c_2^2+c_3^2)^2 + B(c_1^2c_2^2+c_2^2c_3^2+c_3^2c_1^2) \\ \nonumber
& & +B(c_1^2+c_2^2+c_3^2)(c_3^2+c_1^2)+C\;c_3^2c_1^2
\end{eqnarray}
and,
\begin{eqnarray} \label{ThirdEquation}
\lambda & = & 3A (c_1^2+c_2^2+c_3^2)^2 + B(c_1^2c_2^2+c_2^2c_3^2+c_3^2c_1^2) \\ \nonumber
& & +B(c_1^2+c_2^2+c_3^2)(c_1^2+c_2^2)+C\;c_1^2c_2^2
\end{eqnarray}

\ref{FirstEquation}$\times c_1^2$+~\ref{SecondEquation}$\times c_2^2$+~\ref{ThirdEquation}$\times c_3^2$ 
gives 
\begin{eqnarray*} 
\lambda (c_1^2+c_2^2+c_3^2) & = & 3 A (c_1^2+c_2^2+c_3^2)^3 + \\ \nonumber
&& 3 B(c_1^2+c_2^2+c_3^2)(c_1^2c_2^2+c_2^2c_3^2+c_3^2c_1^2)+ \\ \nonumber
&& 3C\;c_1^2c_2^2 
\end{eqnarray*}

This means, the points $(c_1,c_2,c_3)$ which lie on a sphere of radius $k$ and which extremize $I$ satisfy the above equation. But the R.H.S of the above equation is $3I$. If we want $I\geq0$ for all values on the sphere, we need 
to have $\lambda\geq0$. Also, among the extrema points, the point which gives the least value of $I$ (global minimum) also gives the least value of $\lambda$.

Thus we need to find the point associated with global minimum and make the $\lambda$ associated 
with it greater than or equal to zero.

Instead of doing this, we define (a) $\lambda$ consistent with Eq.~\ref{FirstEquation},~\ref{SecondEquation}, and~\ref{ThirdEquation}, (b) find the point 
which gives its global minimum and (c) show that this point also lies on the sphere and 
satisfies Eq.~\ref{FirstEquation},~\ref{SecondEquation}, and~\ref{ThirdEquation}. Thus, it 
is the point associated with the global minimum of $I$ also.

\ref{FirstEquation}+~\ref{SecondEquation}+~\ref{ThirdEquation} gives 
$$ (9A+2B) (c_1^2+c_2^2+c_3^2)^2 + (3B+C)(c_1^2c_2^2+c_2^2c_3^2+c_3^2c_1^2)= 3 \lambda $$

Let us define our $\lambda$ by this equation above. From~\ref{ConstraintForB}, 
we know that $(9A+2B)>0$, if $(3B+C)$ is also greater than or equal 
to zero we have $\lambda>0$ for all $c_i\neq0$. Thus the points extermizing $I$ would satisfy 
the above equation and as $\lambda= 3 I/k^2$ for these points(from Eq.~\ref{ThirdEquation}), 
$I$ would be greater than zero. As the most minimum 
value of $I$ is greater than or equal to zero we have $I\geq$ for all points $(c_1,c_2,c_3)$ on the sphere.
If instead $(3B+C)<0$ the point which gives the global minimum of $\lambda$ is $c_1=c_2=c_3$. One can verify that
this point satisfies  Eq.~\ref{FirstEquation},~\ref{SecondEquation}, and~\ref{ThirdEquation}. Thus,
 it is a point of extremum of $I$, and as $\lambda= 3 I/k^2$ for 
the points of extremization of $I$ we have the global minimum of $I$ also occurring at $c_1=c_2=c_3$. 
And for the $\lambda$ associated with this to be greater than or equal to zero we need
$$-(9A+2B)\leq\frac{3B+C}{3} $$

\begin{equation}
\implies C\geq-9(3A+B) 
\end{equation}

As nowhere in the above derivation, the radius of the sphere to which the points are constrained to, 
played any role in determining the evaluation of the constraints, the result holds for sphere.
Thus, the constraints are evaluated and are given by  Eq.~\ref{FirstEquation},~\ref{SecondEquation}, and~\ref{ThirdEquation}. By connecting how the components $\alpha^I_{ijklmn}$ are related to 
$A$, $B$ and $C$ one can get the constraints on these too (as outlined in Section~\ref{Section5}).

\bibliographystyle{apsrev}
\bibliography{References}

\end{document}